\documentclass[pre,twocolumn,superscriptaddress]{revtex4}
\usepackage{graphicx}
\usepackage{amsmath}
\usepackage{braket}
\usepackage{amssymb,bm}
\usepackage{bm,hyperref}
\usepackage{CJK,color}

\begin{document}
\begin{CJK*}{UTF8}{gbsn}
\title{Roles of energy eigenstates and eigenvalues in equilibration of isolated quantum systems}
\author{Shaoqi Zhu(朱少奇)}
\affiliation{International Center for Quantum Materials, Peking University, 100871, Beijing, China}
\author{Biao Wu(吴飙)}
\affiliation{International Center for Quantum Materials, Peking University, 100871, Beijing, China}
\affiliation{Collaborative Innovation Center of Quantum Matter, Beijing 100871, China}
\affiliation{Wilczek Quantum Center, Department of Physics and Astronomy, Shanghai Jiao Tong University, Shanghai 200240, China}

\date{\today}

\begin{abstract}
We show that  eigen-energies and energy eigenstates  play different roles in the equilibration
process of an isolated quantum system. Their roles are revealed  numerically by exchanging
the eigen-energies between an integrable model and a non-integrable model.
We find that the structure of eigen-energies of a non-integrable model characterized by
non-degeneracy ensures that quantum revival occurs rarely whereas the energy eigenstates
of a non-integrable model suppress the fluctuations for the equilibrated quantum state.
Our study is aided with a quantum entropy that describes how randomly
a wave function is distributed in quantum phase space. We also demonstrate
with this quantum entropy the validity of Berry's conjecture for energy eigenstates. This implies
that the energy eigenstates of a non-integrable model appear indeed ``random".
\end{abstract}
\maketitle
\end{CJK*}

\section{introduction}
How equilibration is achieved in an isolated quantum system is a fundamental
issue regarding the foundation of quantum statistical mechanics. This issue
has intrigued many physicists
~\cite{Neumann,vonNeumann2010qhtheorem,Goldstein1}.  In standard
textbooks on quantum statistical physics, one just assumes that quantum equilibration
can be achieved and assign the equilibrated state certain properties with postulates, such as
equal {\it a priori} probability, to establish micro-canonical ensemble.
In his well-known textbook~\cite{Huang}, Huang states, "The postulates of
quantum statistical mechanics are to be regarded as working hypotheses whose
justification lies in the fact that they lead to results
in agreement with experiments. Such a point of view is not entirely satisfactory,
because these postulates cannot be independent of, and should be derivable from,
the quantum mechanics of molecular systems". This issue has recently received renewed
interests~\cite{Typicality2006prl,Popescu2006Nphys,Reimann,Goldstein2,linden2009thermalequilibrium,Fine2009pre,Ueda2011pre,Yukalov2011lpl,Gogolin2011prl,Eisert2012prl,Short2011njp,Kastner2012njp,Sugiura2012prl,Shimizu2013prl,ZhuangWu,Zhuang,Han,Zhang} due to experimental developments~\cite{Schmiedmayer2013njp,Zhou}

In 1929 von Neumann addressed this fundamental issue by proving
quantum ergodic theorem and quantum $H$ theorem~\cite{Neumann,Goldstein1}, where he claimed, ``in quantum mechanics one can prove the ergodic theorem and the H-theorem in full rigor and without disorder assumptions; thus, the applicability of the statistical-mechanical methods to thermodynamics is guaranteed without relying on any further hypotheses." Now these two theorems have
been re-formulated in a more rigorous framework without invoking ambiguous
``coarse-graining"~\cite{Reimann,Short2011njp,Han}. It is clear in these studies that
the structure of the quantum
system's eigen-energies plays a crucial role: when the eigen-energies and their gaps
are non-degenerate, then the two theorems hold and the isolated quantum system can equilibrate.
The form of energy eigenstates, that is, how they distribute either in position space or in momentum space,
is not important in these studies.  This is, of course, in agreement with the Gibbs distribution at equilibrium which is solely determined by the energies and density of states.
Due to its fundamental role in quantum equilibration,
the structure of eigen-energies was used to give a precise definition of
quantum ergodicity and quantum mixing~\cite{Zhang}.

Recently, a different point of view on quantum equilibration, which was already mentioned
in Landau's book as a footnote~\cite{Landau},  has received a great attention.
This view is eigenstate thermalization hypothesis(ETH), which is justified on the basis of
random matrix theory and Berry's conjecture~\cite{Srednicki,Deutsch,Rigol1}.
According to ETH, the form of energy eigenstates is crucial. In integrable systems,
the eigenstates look rather ``regular"
and are not thermalized; in non-integrable systems, the eigenstates should look ``random"
according to Berry's conjecture and are therefore thermalized. Many numerical and theoretical results~\cite{Rigol1,Rigol2,Rigol3,Ikeda,Kim,Palma,Khodja,Hosur,Borgonovi} on real many-body systems including integrable and non-integrable systems turn out to support this hypothesis and this has stimulated enormous research on many-body localization~\cite{Pal,Mondaini,Nandkishore,Bera,Li1,Vosk,Serbyn}.

Although these two points of view are different, they do not contradict each other.
Most importantly,  they agree on one very important point: only non-integrable isolated
quantum systems can equilibrate or thermalize. In this work we try to clarify
the roles played by eigen-energies and energy eigenstates in quantum equilibration
by comparing an integrable model and a non-integrable model and exchanging their
eigen-energies.

For an isolated quantum system, its dynamics is given by
\begin{equation}
\ket{\psi(t)}=\sum_n a_n e^{-iE_nt/\hbar}\ket{\phi_n}\,,
\end{equation}
\label{dynamic}
where $\ket{\phi_n}$ is an energy eigenstate with eigen-energy $E_n$. The coefficents $a_n$ are
independent of time and determined by the initial state. The dynamics
is clearly controlled by both eigenstates $\ket{\phi_n}$ and eigen-energies $E_n$.
For a given quantum system, if it is integrable, then both $\ket{\phi_n}$ and  $E_n$ show
characteristics of an integrable system; if it is non-integrable, then both $\ket{\phi_n}$ and  $E_n$
are embedded with the features of a non-integrable system. However, numerically,
we can have a dynamics which is controlled by a set of integrable eigen-energies $E_n$
with a set of non-integrable eigenstates $\ket{\phi_n}$. Consider two models,
one is integrable and the other is non-integrable. Suppose their eigenstates
and eigen-energies are, respectively, $\{\ket{\phi_n^i}\,,E_n^i\}$ and $\{\ket{\phi_n^c}\,,E_n^c\}$.
By exchanging the two sets of eigen-energies, we can numerically have four different dynamical
evolutions: 1) integrable eigenstates and integrable eigen-energies; 2) integrable eigenstates and non-integrable eigen-energies; 3) non-integrable eigenstates and integrable eigen-energies; 4) non-integrable eigenstates and non-integrable eigen-energies.  For example, the third dynamics
can be expressed as
\begin{equation}
\ket{\psi'(t)}=\sum_n a_n e^{-iE_n^it/\hbar}\ket{\phi_n^c}\,.
\end{equation}
The second and third dynamics never occurs in a really system. However,
by studying them we are able to clarify the roles played by energy eigenstates and
eigen-energies in dynamics: the non-degeneracy of eigen-energies ensures that
the initial state is dephased over time and  quantum revival is suppressed;
the ``randomness" in the non-integrable  eigenstates keeps the fluctuations around
the equilibrium small.   Therefore, non-degenerate eigen-energies  and "randomized" eigenstates
are equally important for quantum equilibration of a non-integrable system but playing different
roles.

In our numerical study, we use the quantum entropy for pure states introduced in Ref.~\cite{Han}
to quantify the equilibration process. This quantum entropy is defined by projecting
a wave function unitarily to phase space and describes how a wave function is distributed
in phase space.  The more randomly the wave function is distributed the bigger the entropy. In
the second part of our work, we use this entropy to check the validity of Berry's conjecture~\cite{Berry}
and show that the eigenfunctions of a non-integrable system indeed look ``random".
Our study finds that the quantum entropy for energy eigenstates agrees very well with  Berry's conjecture at each energy level and the entropy fluctuation among different eigenstates is very small for the fully chaotic systems. Note that
the validity of Berry's conjecture has been checked previously with  autocorrelation functions~\cite{Shapiro}, amplitude distributions~\cite{Seba,Alt,Kudrolli}, and statistics of nodal domains ~\cite{Blum}.

We organize the paper as follows. In Section II, we briefly describe ripple billiards,  and quantum
phase space, and  the concept of quantum entropy for pure states. In Section III, we compare the time evolution of a Gaussian wave packet moving in a square billiard and a ripple billiard, representing integrable and non-integrable systems, respectively. We then exchange their  eigen-energies to
create two artificial dynamics.  By comparing these different dynamics,
we are able to identify the roles of eigen-energies and
eigenfunctions in quantum equilibration of an isolated system. Section IV is to explain
why eigenfunctions in a chaotic system can play the role identified in the previous section.
This is achieved by comparing them to the wave functions constructed according to
Berry's conjecture. We conclude in Section V.

\section{Model and quantum entropy for pure state}
In this section, we briefly introduce the models for our numerical calculation,
quantum phase space, and the quantum entropy for pure states that we use
to characterize the quantum equilibration. \\
\begin{figure}[htbp]
	\includegraphics[height=5.5cm]{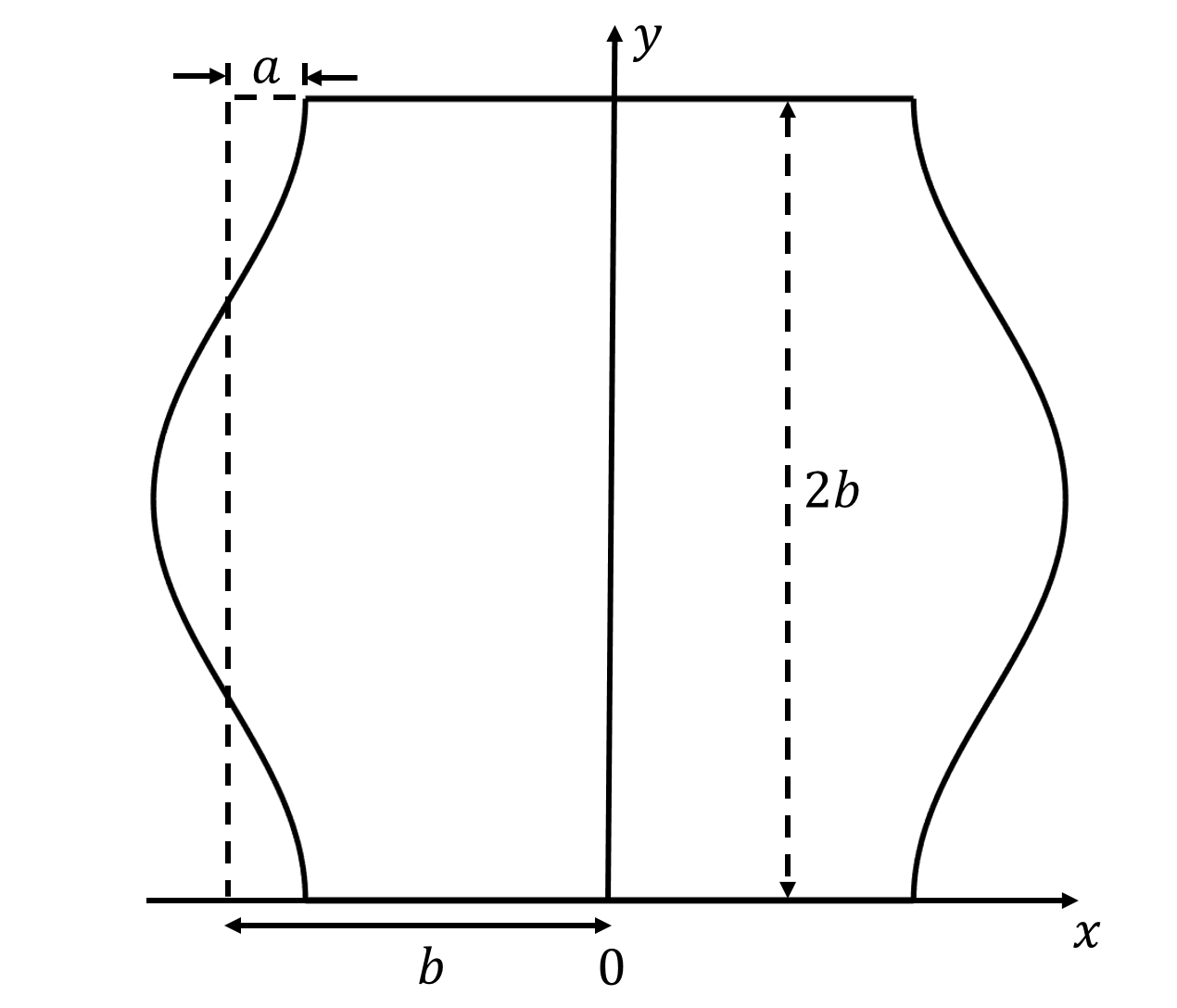}
	\caption{ Ripple billiard. The two curved boundaries are given by
	$x = \pm b \mp a \cos(\pi y/b)$, respectively.}
	\label{billiard}
\end{figure}

In our numerical calculation, we use the model of  ripple billiard~\cite{Li2}, which
is shown in Fig.\ref{billiard}. When $a=0$,  it becomes the square billiard and it is an integrable
system. When $a>0$, it is non-integrable. In general, as $a$ becomes larger, the billiard
is more chaotic~\cite{Li2}. The billiard is special in that the elements of its Hamiltonian
can be calculated analytically. As a result, one can conveniently
study its eigenenergies and eigenstates in a systematic way. Details can be found
in Ref. ~\cite{Li2}.\\

\begin{figure}[!htbp]
	\includegraphics[width=7cm]{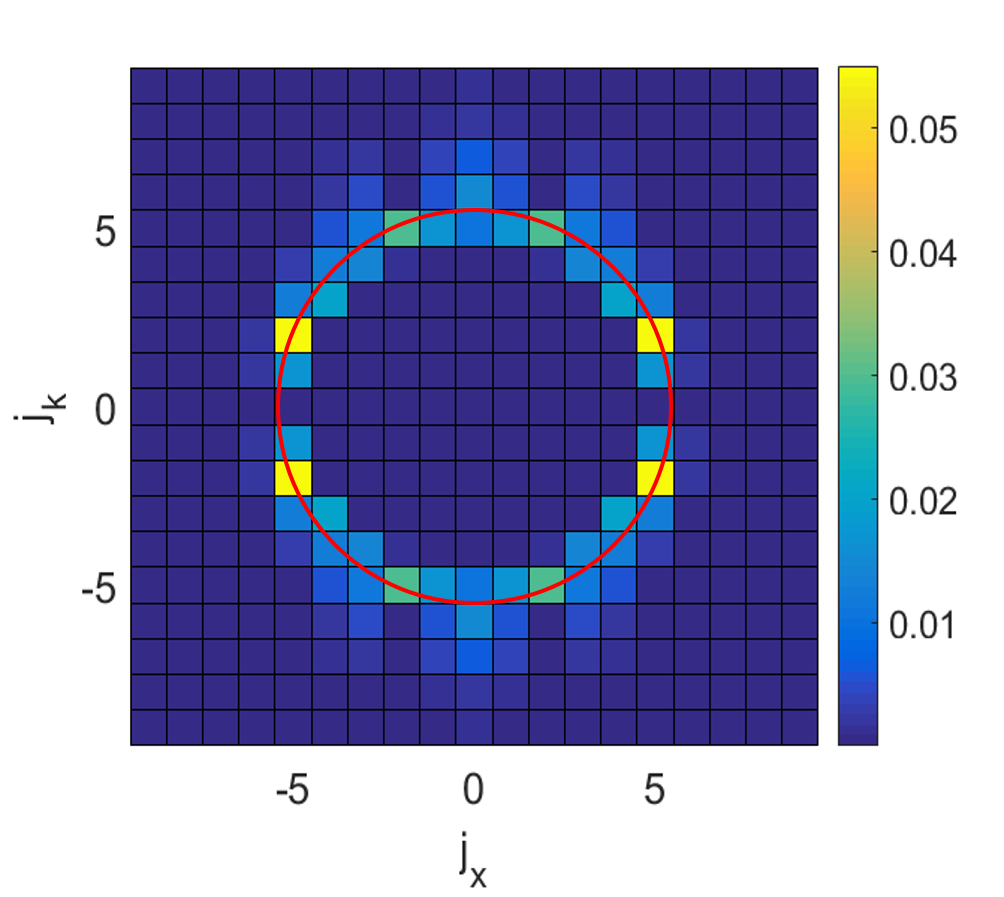}
	\caption{(color online) The $100$th eigenfunction of a one-dimensional harmonic oscillator
	in the quantum phase space. The red circle is the corresponding classical trajectory. $j_x$ and $j_k$
	are indices labelling Planck cells.}
	\label{ho}
\end{figure}

Besides the well known von Neumann entropy, another quantum entropy was introduced by
von Neumann in his 1929 paper~\cite{Neumann}. This quantum entropy was defined
for pure states. However, von Neumann's definition involves ambiguous coarse-graining, making
numerical computation impossible. In Ref. \cite{Han}, von Neumann's definition
was modified and a new quantum entropy for pure states was defined with
Wannier functions obtained with Kohn's method~\cite{Kohn}. To define this entropy,
we need first to construct a quantum phase
space:  (1) the classical phase space is divided into Planck cells; (2) each
Planck cell is assigned a Wannier function and all the Wannier functions form a set of a
complete orthonormal basis~\cite{Han}. The Wannier functions
are constructed by orthonormalizing  a set of  Gaussian wave packets of width $\zeta$,
\begin{equation}
g_{j_x,j_k}\equiv \exp\left[-\frac{(x-j_xx_0)^2}{4\zeta^2}+ij_kk_0x\right],
\end{equation}
where $j_x$ and $j_k$ are integers. When $x_0k_0=2\pi$, this set of the resulted Wannier functions  is complete. In this paper, parameters are chosen as $x_0=1, k_0=2\pi$, and $\zeta=(2\pi)^{-1}$.
The details of this construction of quantum phase space can be found in Ref. \cite{Han}.
Once the Wannier functions are obtained, they are used to project a wave function unitarily onto the quantum phase space. To give unfamiliar  readers a sense of this quantum phase space,  the $100$th eigenfunction of a one-dimensional harmonic oscillator is mapped in this quantum phase space and is shown in Fig. \ref{ho}.
The wave function concentrates on the classical trajectory.

If $\ket{w_j}$ is the Wannier function at Planck cell $j$, then $|\langle\psi|w_j\rangle|^2$ is the probability at Planck cell $j$ for a wave function $\psi$. Our quantum entropy for  pure state $\psi$
is defined  with these probabilities as
\begin{equation}
S_w(\psi)\equiv-\sum_j\langle\psi|\bm{W}_j|\psi\rangle \ln\langle\psi|\bm{W}_j|\psi\rangle,
\label{entropy}
\end{equation}
where $\bm{W}_j\equiv|w_j\rangle\langle w_j|$ is the projection to Planck cell $j$. It is clear
from this definition that the entropy $S_w(\psi)$ describes how a quantum state $\psi$
is spread out in the phase space: the more Planck cells that $\psi$ occupies the bigger its entropy.

In our numerical calculation, length is  in an arbitrary unit of $L$. Correspondingly,
the wave vector $k$ is in unit of $1/L$ and the energy is in unit of $\hbar^2/2mL^2$, where $m$
is the particle mass.  Throughout this paper we omit these units for convenience. For example,
when we say $b=5.5$ we mean $b=5.5L$.  The $j$ in $w_j$ stands for $\{j_x,j_{k}\}$  in a one dimensional system and   $\{j_x,j_y,j_{k_x},j_{k_y}\}$ in a two dimensional system.

Here are the details on  the quantum phase space in our numerical calculation. Taking $b=5.5,\ a/b=0.2$ for example, the ripple  billiard is confined in a rectangle area $13.2\times11$.
Every Planck cell in position space is $1\times1$.
When we map a wave function in a ripple billiard onto the phase space, we need $N_{j_{\bm{x}}}=13\times11$  position indices with $j_x\in[-6,6]_\mathbb{Z}$ and $j_y\in[1,11]_\mathbb{Z}$ to cover the whole real space.  The maximum wavelength corresponding to the energy scale in  our numerical computation is $|\bm{k}|=4\times2\pi$. Therefore, we need $N_{j_{\bm{k}}}=9\times9$ momentum indices with $j_{k}=[-4,4]_\mathbb{Z}$ in both the $k_x$ direction and the $k_y$ direction.
The total number of Planck cells is $N=N_{j_{\bm{x}}}\times N_{j_{\bm{k}}}$. If the wave function $\psi$ distributes equally in the $N$ Planck cells, the entropy would be $S_{max}=\ln N$. The mesh points is $180\times180$ dividing the billiard into numerically discrete area.

\section{Time evolution}
Our main aim of this work is to identify  the roles played by eigenstates and eigen-energies
in quantum dynamics, particularly in the dynamics that leads to equilibration.  For this purpose,
we choose two different billiards: (1) square billiard ($a/b=0$); (2) chaotic ripple billiard ($a/b=0.2$).
We not only study and compare their dynamics but also create two artificial dynamics by
exchanging these two billiards' eigen-energies. Let
$\{\ket{\phi_n^i}\,,E_n^i\}$ be the set of eigenstates and eigen-energies for the square billiard
and $\{\ket{\phi_n^c}\,,E_n^c\}$ be the set of eigenstates and eigen-energies for the ripple billiard.
The dynamics of these two billiards can be described formally as
\begin{equation}
\ket{\psi_i(t)}=\sum_n a_n e^{-iE_n^it/\hbar}\ket{\phi_n^i}\,,
\label{di}
\end{equation}
and
\begin{equation}
\ket{\psi_c(t)}=\sum_n b_n e^{-iE_n^ct/\hbar}\ket{\phi_n^c}\,,
\label{dc}
\end{equation}
where the coefficients $a_n$'s and $b_n$'s are determined by the initial condition.
By exchanging their eigen-energies, we can create two more dynamics
\begin{equation}
\ket{\psi_{ic}(t)}=\sum_n a_n e^{-iE_n^c t/\hbar}\ket{\phi_n^i}\,,
\label{dic}
\end{equation}
and
\begin{equation}
\ket{\psi_{ci}(t)}=\sum_n b_n e^{-iE_n^it/\hbar}\ket{\phi_n^c}\,.
\label{dci}
\end{equation}
These two dynamics are artificial but will help us to identify the roles of eigenstates and
eigen-energies.

\begin{figure}[!htbp]
	\includegraphics[width=6.3cm]{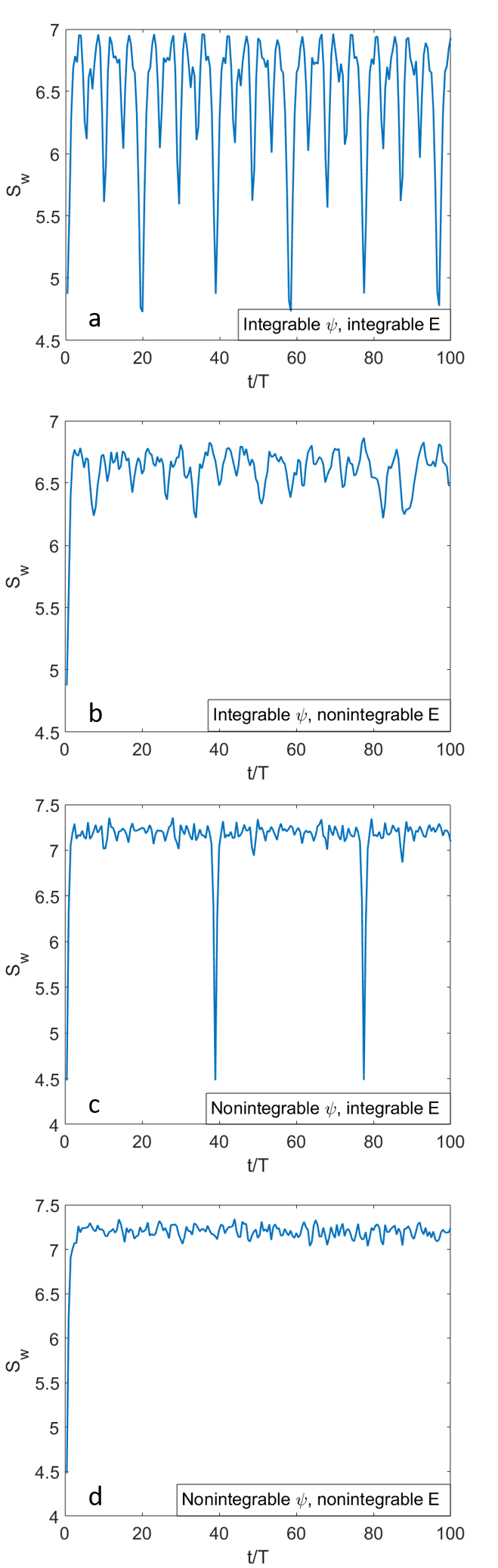}
	\caption{(color online) Time evolution for the quantum entropy in four situations: (a) integrable eigenstates and integrable eigenenergies; (b) integrable eigenstates and nonintegrable eigenenergies; (c) nonintegrable eigenstates and integrable eigenenergies; (d) nonintegrable eigenstates and nonintegrable eigenenergies. The scale of the ripple billiard is $a=0.55, b=5.5$ while the length of side for the square billiard is $b$. Initial Gaussian wave packet parameters: $\sigma=1$, $k_y=0$; $k_x=6.05$ for ripple billiard and $k_x=5.5$ for square billiard. $T$ is the characteristic time when the center of the initial wave packet return to the center of the billiard after reflecting once from two boundaries along $x$ direction. $L$ is the length unit; $1/L$ is the unit for $k$. }
	\label{time_evolution}
\end{figure}

The numerical results of these four dynamics are shown in Fig.\ref{time_evolution}.
The initial state for these four different dynamics is the same and it is a moving Gaussian
wave packet,
\begin{equation}
\Psi(0)=\exp\left[-\frac{x^2+(y-b)^2}{4\sigma^2} + i(k_xx+k_yy)\right]\,.
\end{equation}
With numerically computed eigen-fucntions
$\ket{\phi_n^i}$ and $\ket{\phi_n^c}$, we determine the coefficients $a_n$'s and $b_n$'s.
This allows us to find the wave functions at any time $t$. We finally compute
the entropies for these wave functions with Eq.(\ref{entropy}). How
the entropies change with time is shown in Fig.\ref{time_evolution}.

There are four very different dynamics in Fig.\ref{time_evolution}. In case (a) (integrable
eigenstates and integrable eigenenergies), the entropy oscillates regularly with time with large amplitudes. In case (b) (integrable eigenstates and nonintegrable eigenenergies), the entropy
increases quickly to a large value and stay at this value with relatively large fluctuations.
In case (c) (nonintegrable eigenstates and integrable eigenenergies), the entropy similarly
relaxes quickly to a large value with small fluctuations. However, the entropy drops
back almost to its initial value after a certain period. This period is consistent with
the oscillation period in the case (a). This is the well known phenomenon of quantum revival.
In case (d) (nonintegrable eigenstates and nonintegrable eigenenergies),
 the entropy quickly relaxes to its maximum value and stays there with very small fluctuations.
There is no quantum revival.

The results in Fig.\ref{time_evolution} are quite revealing. To reach quantum
equilibrium as in Fig.\ref{time_evolution} (d),  we need both nonintegrable eigenstates and nonintegrable eigen-energies. The nonintegrable eigenstates ensure that the fluctuations
are small once the equilibrium is reached. The nonintegrable eigen-energies are a must
for  no occurence of large deviation in a physically meaningful time.
These two important points are not hard to understand intuitively:
the nonintegrable eigen-energies lack of degeneracy in eigen-energies and
their gaps that is needed for regular quantum dynamics;  the nonintegrable eigenstates
are rather ``random" according to Berry's conjecture;
The former has been discussed extensively in Ref.\cite{Neumann,Reimann,Han,Zhang}.
We will examine the latter  in detail in the next section.

 \begin{figure*}[htbp]
	\includegraphics[width=17cm]{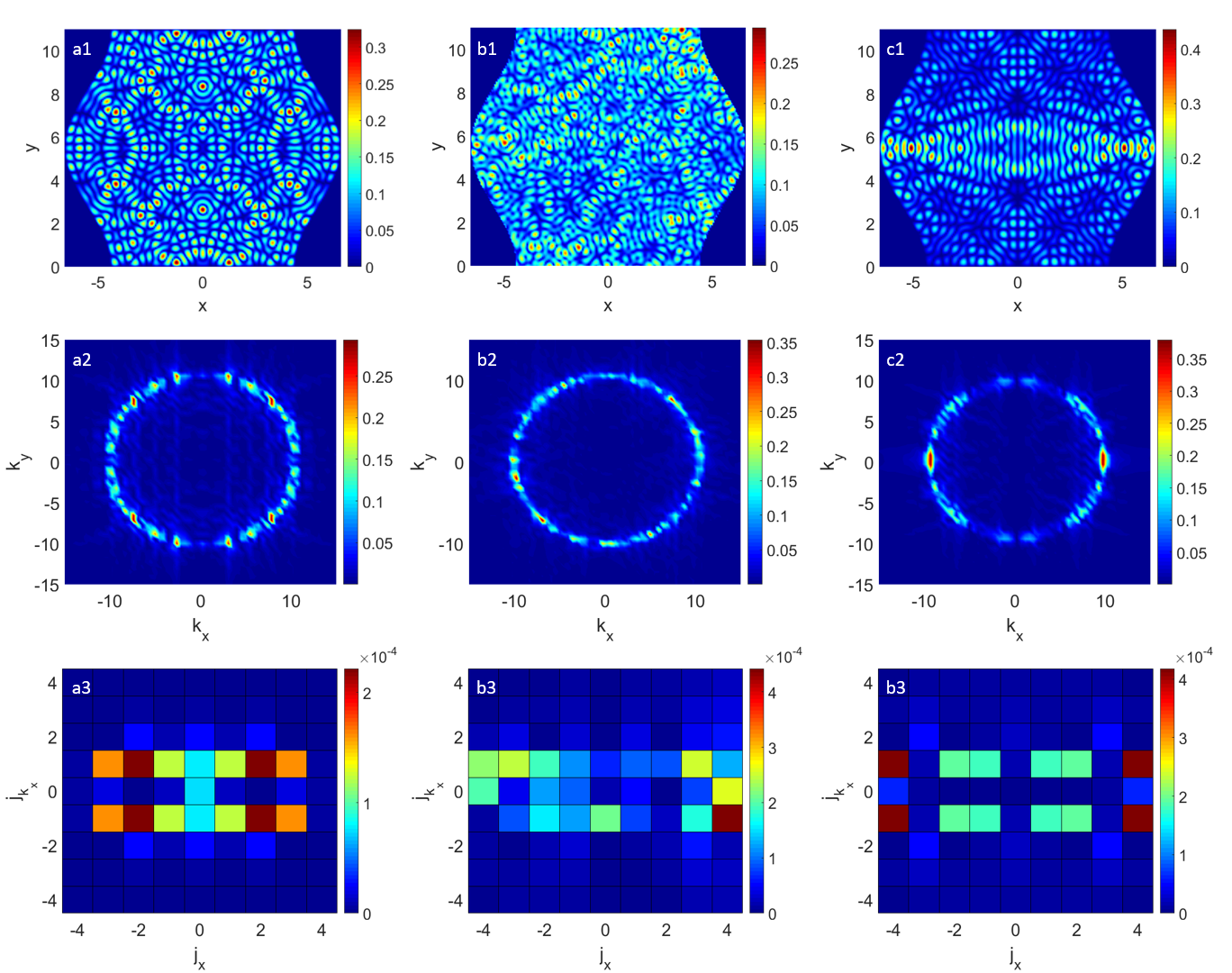}
	\caption{(color online) (a1) the 1000th eigenstate in the position space; (b1) the corresponding wave function $\psi_B$ constructed according to Berry's conjecture in the position space; (c1)
	the $857$th eigenstate (which is a scar state) in the position space. (a2,b2,c2) Their
	respective representation in the momentum space. (a3,b3,c3) Their representations in
	the quantum phase space with  the position  and momentum along the $y$ direction
	fixed at $j_y=5, j_{k_y}=0$. For the billiard, $a=0.55$ and $b = 5.5$. $L$ is the unit of length.  }
	\label{wavefunction}
\end{figure*}

\section{ Entropy for Eigenstates and Berry's conjecture}
In the last section, we see that nonintegrable eigenstates are essential to keep the
fluctuations small at equilibrium.  The intuitive reason is that these
nonintegrable eigenstates are ``random" according to Berry's conjecture, which is
the base for ETH~\cite{Srednicki}. However, there are two important issues that
have so far no satisfactory answers. The first one is how to measure quantitatively
the  ``randomness'' in eigenstates. If there is such a measure of randomness,
how the eigen-wavefunction constructed artificially according to Berry's conjecture
compares to the real eigenstates? The other issue is
that there are many quantum scar eigenstates. These eigenstates look regular as
their amplitudes concentrate along classical periodical orbits~\cite{Heller}. How often
do they appear? If  there is  a quantitative measure of randomness, how far these
quantum scar states deviate from other eigenstates? We examine these two issues
in this section.

The quantum entropy $S_w(\psi)$ defined in Eq.(\ref{entropy}) is a good measure
of the randomness in eigen-wavefunctions. As the wave function is project
unitarily onto the quantum phase space, $S_w(\psi)$ contains information both in position and
momentum. In contrast, the probability $\psi(x)$ ($\psi(k)$) has information only in
position (or momentum). We shall use it to measure the randomness in eigen-wavefunctions.

Berry's conjecture states that each eigenfunction of a classically chaotic quantum billiard system is a superposition of plane waves with random phase and Gaussian random amplitude but with the same wavelength\cite{Berry, Srednicki}. Mathematically, such a wavefunction with wave length $k$ can be expressed as
\begin{equation}
\psi_B = \int d\bm{k} A(\bm{k})\exp\{-i\left[\bm{k}\cdot\bm{r} + \theta(\bm{k})\right]\},
\end{equation}
where the modulus of $\bm{k}$ is fixed but it can point to any direction.
Amplitude $A(\bm{k})$ is a Gaussian random distribution for $k$ in different
direction. $\theta(\bm{k})$ is the random phase.

\begin{figure*}[htbp]
	\includegraphics[height=9cm]{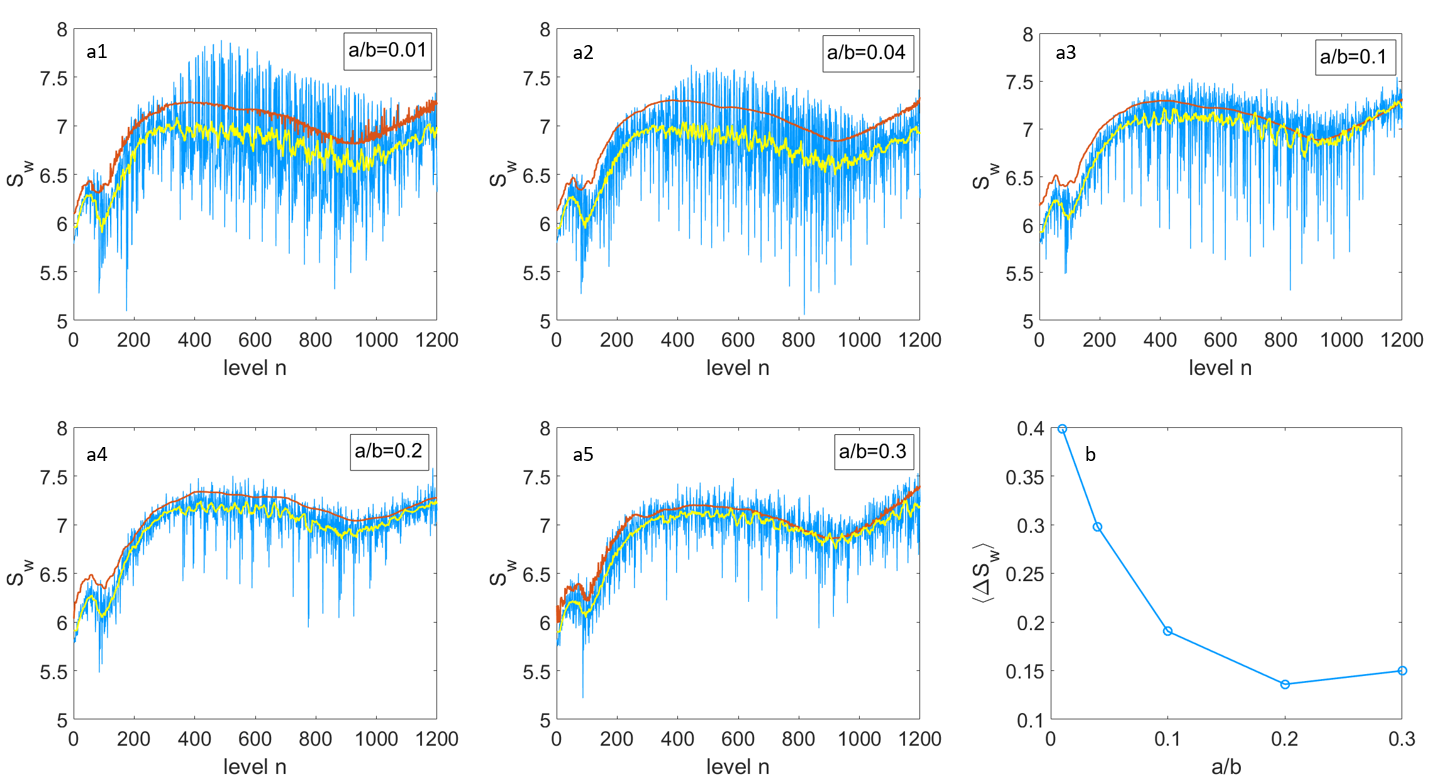}
	\caption{(color online)  (a1-a5) Quantum pure state entropy $S_w$ of eigenfunctions of the ripple
	billiards and their corresponding $\psi_B$ constructed according to Berry's conjecture.
	 The $x$ axis is the eigenenergy level. Blue lines represent the results of the ripple billiard;
	 red lines represent the results for $\psi_B$.  Yellow lines are obtained from the blue lines by  averaging the nearest 30 eigenstates (the standard micro-canonical ensemble average).
	 (b) Average entropy fluctuation around its micro-canonical ensemble average for different $a/b$.}
	\label{Two_entropy}
\end{figure*}

For comparison, we calculate the wave functions $\psi_B$ for every wavelength $k$ that
corresponds to an eigenstate of the ripple billiards. We first look at an example,
where $\psi_B$ is computed with the wavelength corresponding to
the $1000$th eigenstate for the ripple billiard ($a=0.55,~b=5.5$). These two wave functions
are plotted  in Fig.\ref{wavefunction} with the $857$th eigenstate, which is a scar state~\cite{Heller}.
The wave functions are compared in three different spaces: in position space, in momentum space,
and in quantum phase space.  It is clear from the figure that the $1000$th eigenstate
and its corresponding $\psi_B$ are qualitatively similar: both their wave functions are
quite spread-out in all these three spaces. This is confirmed by our entropy:
for the $1000$th eigenstate $S_w=7.11$; for the Berry wavefunction $\psi_B$,
$S_w=7.09$.  As a scar state, the $857$th eigenstate looks qualitatively different from
the $1000$th eigenstate. In the position space, the $857$th eigenstate concentrates on a periodic
trajectory that describes a classical particle bouncing horizontally in the middle of the
billiard. As a result, its momentum distribution concentrates along certain directions and
its distribution in the phase space focuses on some areas. Quantitatively, its
entropy is $S_w=6.57$, significantly smaller than the other two wave functions.
Note that $\psi_B$ is constructed without respecting the symmetry of the system so that it
does not have the symmetries that we see in the $1000$th and $857$th eigenstate.

We now compare the Berry wave functions $\psi_B$ and the  eigenstates of ripple billiards systematically. For a given billiard, the entropies are computed for its eigenstates  from the $1$st to $1200$th and their corresponding Berry wave functions $\psi_B$. The results
for five different billiards are shown in Fig. \ref{Two_entropy}, where the blue lines
are for the eigenstates and the red lines are for $\psi_B$.  For the billiard with $a/b=0.01$,
we see that the entropies of eigenstates have a general trend to increase with energy levels
and this trend is shared by the Berry wave functions $\psi_B$. However, the entropies of eigenstates
have much larger fluctuations compared to $\psi_B$.  As we increase the ratio $a/b$ and
the billiard gets more chaotic ~\cite{Li2,Zhuang}, the general trend of the entropy does not
change. However, the fluctuations become smaller.  This is quantitatively shown in the last panel.
Our numerical observation is that the large fluctuations for the billiards with $a/b\ge 0.1$ are caused
by scar states which is about 10\% of all the eigenstates. Note that for the billiards with small $a/b$,
they are near integrable and it is hard to distinguish scar states and other regular-looking eigenstates.

We have also averaged the entropy over every nearest 30 eigenstates. The results are plotted as
yellow lines in Fig. \ref{Two_entropy}. Even for near-integrable billiards,
the averaged entropy agrees well with the entropy of the Berry wave function $\psi_B$ with
small fluctuations. The agreement improves as the ratio $a/b$ increases. Such an agreement
implies that once the averaging is over a large number of eigenstates how each eigenstate
looks is no longer important. This shows that the postulate of equal {\it a priori} probability in standard textbook~\cite{Huang} over an energy shell of many eigenstates is
valid even for integrable systems. That is why it is not necessary to discuss
integrability of a system in standard textbooks on quantum statistical
mechanics~\cite{Huang,Landau}.

\section{conclusion}
We have  identified the roles of eigenstates and eigenenergies in quantum equilibration of
an isolated system. This is achieved by exchanging the set of eigen-energies
between an integrable system and a chaotic system in our numerical simulations.
Both  the non-degeneracy of eigen-energies and the ``randomness" in eigenstates are equally important for a non-integrable system to achieve equilibration. The non-degeneracy of eigen-energies ensures the initial state is dephased over time and that the  quantum revival is suppressed.
The ``randomness" in the non-integrable  eigenstates keeps the fluctuations around the equilibrium small.
We have also shown in terms of a  quantum pure state entropy that Berry's conjecture
can quantitatively captures the "randomness" of the eigenstates.

\section{acknowledgement}
We thank Xizhi Han for helpful discussions and Dongliang Zhang for his codes on triangular billiard.
This work was supported by the National Basic Research Program of China (Grants No. 2013CB921903) and the National Natural Science Foundation of China (Grants Nos. 11334001 and 11429402).

\bibliographystyle{apsrev4-1}
\bibliography{reference}

\end{document}